\documentclass[aps,prl,reprint]{revtex4-1}
\usepackage{amsmath}
\usepackage{amssymb}
\usepackage{pxfonts}
\usepackage{graphicx}
\usepackage{eucal}
\usepackage{mathrsfs}
\usepackage{theorem}
\usepackage{pifont}
\usepackage{color}
\usepackage{subfig}
\usepackage{cancel}
\newcommand{\brk}[1]{\left( #1 \right)}

\newcommand{\pd}[2]{\frac{\partial #1}{\partial #2}}

\newcommand{\matrixII}[4]{\left(\begin{array}{cc}#1&#2\\#3&#4\end{array}\right)}

\newcommand{\e}{\varepsilon}
\newcommand{\vp}{\varphi}

\newcommand{\figref}[1]{Fig.~\ref{#1}}
\newcommand{\tabref}[1]{Table~\ref{#1}}

\newcommand{\go}{\bar{\mathfrak{g}}}

\newcommand{\g}{\mathfrak{g}}
\newcommand{\euc}{\mathfrak{h}}
\newcommand{\A}{\mathcal{A}}
\newcommand{\B}{\mathcal{B}}

\newcommand{\xvec}{\mathbf{x}}
\newcommand{\Qvec}{\mathbf{Q}}
\newcommand{\Dvec}{\mathbf{D}}
\newcommand{\rvec}{\mathbf{r}}
\newcommand{\pvec}{\mathbf{p}}
\newcommand{\dvec}{\mathbf{d}}

\newcommand{\beq}{\begin{equation}}
\newcommand{\eeq}{\end{equation}}

\newcommand{\Exp}[1]{e^{#1}}

\newcommand{\Cof}[1]{\text{Cof}(#1)}

\newcommand{\Kbar}{\bar{K}_G}

\newcommand{\nablago}{\bar{\nabla}}

\begin{document}


\title{Elastic interactions between 2D geometric defects}


\author{Michael Moshe}
\email[]{mmoshe@syr.edu}
\affiliation{Department of Physics, Syracuse University, Syracuse, NY 13244-1130, USA}
\affiliation{Department of Physics, Harvard University, Cambridge, Massachusetts 02138, USA}
\author{Eran Sharon}
\email[]{erans@mail.huji.ac.il}
\affiliation{Racah Institute of Physics, The Hebrew University, Jerusalem}
\author{Raz Kupferman}
\affiliation{Einstein Institute of Mathematics, The Hebrew University, Jerusalem}


\date{\today}

\begin{abstract}
In this paper, we introduce a methodology applicable to a wide range of localized two-dimensional sources of stress. 
This methodology is based on a geometric formulation of elasticity. Localized sources of stress are viewed as singular defects---point charges of the curvature associated with a reference metric.
The stress field in the presence of defects can be solved using a scalar stress function that generalizes the classical Airy stress function to the case of materials with nontrivial geometry. This approach allows the calculation of interaction energies between various types of defects.
We apply our methodology to two physical systems: shear-induced failure of amorphous materials and the mechanical interaction between contracting cells. 
\end{abstract}


\maketitle

\section{Introduction}
A main theme in physics and material science is to predict a material's mechanical properties given its microscopic structure. The study of crystalline materials has a longstanding history, and a fairly good understanding of their mechanics has been obtained. When it comes to amorphous materials the situation is different. There is no fundamental theory, to date, that relates microscopic structure to mechanical behavior in amorphous materials.  

A difference between crystalline and amorphous solids is  their response to external loads. 
When a crystalline solid is weakly loaded, it responds elastically, maintaining its microscopic order. Above a critical load, defects, which are microscopic deviations from the ordered state, may form and move. The formation, movement, and interactions of defects in crystals determine their macroscopic plastic behavior (see for example \cite{Taylor1}).
In amorphous solids, in which no structural order exists, the response to weak loads is also elastic. For larger loads, bonds between neighboring particles may break and reform in small bounded regions, a phenomenon known as localized plastic deformations (LPDs).

Despite the fundamental differences between crystalline and amorphous materials, both react to strong loads similarly---by a local rearrangement of particle bonds. However, as there is no underlying structural order in amorphous materials, it is not clear how to characterize localized deformations, nor how to quantify their interactions.

In a recent work, we advocated that defects in crystals and LPDs in amorphous materials can be described in a unified manner \cite{Moshe2015}. In a continuum theory, in which the material is modeled by a manifold endowed with a reference metric, both defects in crystals and LPDs in amorphous materials can be described as singularities in the reference curvature, associated with the reference metric. The reference curvature is a geometric invariant, which can be calculated directly from the metric. Viewed this way, LPDs can be called defects, even though they are not related to a deviation from an ordered state. Henceforth, the term defect will be used to describe metric singularities in both crystalline and amorphous materials.

It was recently suggested that elastic interactions between defects lay at the heart of plasticity theory in amorphous materials \cite{Procaccia2}. While interactions between defects have been studied extensively in crystalline materials, a unified treatment of the various types of defects is still lacking. E.g., interactions between dislocations \cite{Nabarro1952} are treated differently than interactions between point defects and dislocations \cite{Michel1980}. 
In amorphous materials, there exists a limited amount of work on interaction between localized deformations \cite{Procaccia2}.
Our description of defects in crystalline and amorphous materials as geometric singularities paves the way to a unified approach to defect interactions in both types of materials.
   
To derive the interaction energy between defects, one needs to know the material's state-of-stress in the presence of those defects. This elastic problem is complicated by the geometric nonlinearity induced by the metric singularities. 
In this work we calculate the interaction energy between various types of defects using an approximation method applicable to general metrically-incompatible systems \cite{ISF}. We focus on defects for which the induced 
state-of-stress is quasi two-dimensional, that is, the system is translationally symmetric along an axis. This includes edge-dislocations, Stone-Wales defects, vacancies and inclusions (screw dislocations, for example, remain out of the scope of the present work). 
These defects can either be considered as point singularities on a flattened thin sheet, or as straight line singularities in a 3D elastic medium.

The state-of-stress of a metrically-incompatible material depends, of course, on the material's constitutive properties.
For a homogeneous and isotropic Hookean solid, we obtain analytical expressions for the interaction energy between various types of defects. Our results are in 
excellent agreement with known results for interactions between dislocations, and provide new predictions for interactions between other types of defects. Our approach is not limited to a particular constitutive law, and it can be extended to higher-orders of accuracy.

Finally, we apply our methodology to two physical problems. 
The first problem is related with the failure of amorphous materials under external loads. In particular, we interpret the avalanche behavior observed in failure experiment. The second problem studies the mechanical interaction between contracting cells adhering to an elastic substrate \cite{Balaban2001,Kineret2013,Safran2013}. When a cell contracts, it generates a localized source of stress in the substrate, very much like an LPD in an amorphous material. In this system, the interaction between remote cells is generated by a similar mechanism as interaction between defects in solids. 

\section{Metric formulation of elasticity}

Geometric formulations of elasticity model an elastic body as a Riemannian manifold $\B$ equipped with a reference metric  $\go$ \cite{EfiJMPS}. The reference metric encodes local equilibrium distances between adjacent material elements.
A configuration of an elastic body is an embedding of $\B$ in the ambient Euclidean space. Every configuration induces on $\B$ a metric, $\g$, which quantifies actual distances between adjacent material elements (in the literature $\g$ is known as the right Cauchy-Green tensor). The most common definition of the elastic strain tensor is the discrepancy between the actual metric and the reference metric,
\begin{equation}
u = \frac{1}{2} \brk{\g - \go}
\label{eq:strain}
\end{equation}
(see \cite{Neff2015} for a review on alternative measures of strain).
In particular, in a strain-free configuration, the actual metric and the reference metric coincide everywhere.
Note that this notion of strain is purely geometric and involves no linearization.

In classical elasticity, bodies are assumed strain-free in the absence of external constraints. This statement is equivalent to saying that $\go$ is Euclidean. In many cases of interest, however, the reference metric is non-Euclidean, leading to a theory of incompatible elasticity. 

Incompatible elasticity was first introduced decades ago by Kondo \cite{Kondo}, Bilby \cite{Bilby}, Kr\"oner \cite{kroner} and Wang \cite{Wang1968} in the context of defects in crystalline solids. 
In the aforementioned literature, incompatibility was associated with a non-Riemannian notion of parallelism. In our approach, incompatibility is a purely metric notion that reflects the inability to embed the material manifold isometrically in Euclidean space.

The elastic model is fully determined by a constitutive relation, which relates the internal stresses to the strain field. In the case of a hyper-elastic material, the constitutive relation can be derived from an energy functional, which is an additive measure of a local energetic cost of deviations of the actual metric from the reference metric. In the case of an amorphous solid, we assume that the microscopic structure of the solid is fully encoded by the reference metric $\go$, in which case the elastic energy is of the form,
\begin{equation}
E = \int_\B W(\g(\xvec); \go(\xvec))\, d\text{Vol}_{\go}\,\,\, ,
\label{EnergyForm}
\end{equation}
where $d\text{Vol}_{\go}$ is the Riemannian volume element, and $W$ is a non-negative energy density,  which vanishes at $\xvec$ if and only if $\g(\xvec) = \go(\xvec)$. 
Given a reference metric $\go$, and a specific form of the energy density $W$, the actual metric $\g$ at equilibrium is the one minimizing \eqref{EnergyForm}.

Incompatibility manifests in that $\g$ cannot be equal to $\go$ everywhere simultaneously. 
Incompatibility occurs when the reference curvature of $\go$ is non-zero. 
In two-dimensional systems, the reference curvature is determined by a scalar field, the Gaussian curvature $\Kbar$.
Thus, $\Kbar$ can be viewed as a source of residual stresses. This observation is the key to the description of localized defects by a metric structure in which $\Kbar$ is everywhere zero, except in the loci of the defects.

\section{Metric description of 2D defects}
 
Two-dimensional reference metrics can be written using isothermal coordinates \cite{doCarmoBaby}, 
\begin{equation}
\go = e^{2 \vp(\xvec)} \euc,
\label{eq:Conformal}
\end{equation}
where $\euc$ is the Euclidean metric and $\vp$ is called the conformal factor. The relation between the conformal factor $\vp$ and the reference Gaussian curvature is given by Liouville's equation \cite{liouville1853equation}, 
\begin{equation}
\Kbar = 
-\Exp{-2\vp(\xvec)}\Delta \vp(\xvec).
\end{equation}
Localized defects are hence represented by a metric of the form \eqref{eq:Conformal}, with a conformal factor $\vp$ that is harmonic everywhere except at the loci of the defects, where $\vp$ is singular. 

Consider first a single defect located at the origin. Harmonic functions with singularities at the origin can be expanded in a multipole expansion. The monopole term, 
\[
\vp_\text{M}(\xvec) = \frac{\alpha}{2\pi} \ln  |\xvec|,
\] 
corresponds to a singular reference curvature of the form 
\[
\bar{K}_\text{M} = \alpha \, \Exp{-2\vp_\text{M}(\xvec)} \, \delta(\xvec),
\]
and represents a disclination. The dipole term,
\[
\vp_\text{D}(\xvec) = \frac{\pvec\cdot \xvec}{2\pi |\xvec|^2},
\]  
where $\pvec$ is a vector, corresponds to a singular reference curvature of the form
\[
\bar{K}_\text{D}(\xvec) = \Exp{-2\vp_\text{D}(\xvec)}\,\pvec \cdot \vec{\nabla} \delta(\xvec),
\] 
and represents a dislocation. 

There is a fundamental difference between defects represented by monopole and dipoles terms, and defects represented by higher multipoles (e.g. Eshelby inclusions or Stone-Wales defects). In the latter case,  if the locus of the defect is removed (thus, creating a void), the material can relax to a strain-free configuration. In the former case, the defect is topological in the sense that it persists even if its locus is removed. Thus, higher-order multipoles can be generated by local plastic deformations, whereas the formation of disclinations and dislocations is necessarily nonlocal.

A useful property of the metric description of defects using a conformal factor is the ability to account easily for multiple localized defects. Any locally-Euclidean metric with $N$ singular defects can be represented by a metric \eqref{eq:Conformal}, with a conformal factor of the form,
\beq
\vp(\xvec) = \sum_{i=1}^N \vp_i(\xvec - \xvec_i),
\label{eq:mult_phi}
\eeq
where $\xvec_i$ is the coordinate of the $i$-th defect and $\vp_i$ is a harmonic function singular at the origin.  

This incompatible elasticity model is valid for any material whose elastic state is well-approximated by metric quantifiers, whether it is crystalline or amorphous. When it comes to singular defects, the modeling of the material's geometry by a locally-Euclidean metric is only valid up to a cutoff distance related to the core of the defect. As a result, our analysis of defect interaction is only valid as long as the distance between defects is significantly larger then the size of the cores.

\section{The incompatible stress function} 

The Euler-Lagrange equations obtained by minimizing the elastic energy functional \eqref{EnergyForm} are 
\begin{equation}
\nablago_{\mu} \sigma^{\mu\nu} + \brk{\Gamma^\nu_{\alpha \beta}-\bar{\Gamma}^\nu_{\alpha \beta}} \sigma^{\alpha \beta} = 0,
\label{DivSigma}
\end{equation}
where 
\begin{equation}
\sigma^{\mu\nu} = \pd{W(\g; \go)}{\epsilon_{\mu\nu}} = 2  \pd{W(\g; \go)}{\g_{\mu\nu}}.
\label{eq:sigma}
\end{equation}
is the stress tensor.
$\Gamma$ and $\bar{\Gamma}$ are the Christoffel symbols \cite{docarmo} associated with $\g$ and $\go$, respectively.  The  operator $\nablago$ is the covariant derivative  with respect to $\go$,
namely,
\[
\nablago_{\mu} \sigma^{\mu\nu} =
\partial_\mu  \sigma^{\mu\nu} + \bar{\Gamma}^\mu_{\mu \beta} \sigma^{\beta\nu}
+ \bar{\Gamma}^\nu_{\mu \beta} \sigma^{\beta\mu}.
\]
For traction boundary conditions,
\begin{equation}
n_{\alpha} \sigma^{\alpha \beta} = t^{\beta},
\label{eq:BC}
\end{equation}
where $\mathbf{t}$ is the traction vector and $\mathbf{n}$ is the unit normal to the boundary.

Equation \eqref{DivSigma} is a momentum balance law, and as such is
independent of the material's constitutive law. The latter enters in the relation \eqref{eq:sigma} between the stress and the configuration. The equilibrium equations \eqref{DivSigma}, together with the constitutive law \eqref{eq:sigma} and the
boundary conditions \eqref{eq:BC},  form a closed system of equations.

The dependent variable whose solution is sought is conventionally taken to be the configuration. We adopt a different approach, and express the elastic problem as a system of equations in which the unknown is the actual metric $\g$. The actual metric determines the configuration modulo rigid transformation; thus, such a change of variables is legitimate as long as the boundary conditions do not depend on position.

Any two-dimensional divergence-free tensor field can be expressed as the curl of the gradient of a scalar function. Equation \eqref{DivSigma}, like any momentum balance law, states that the stress is divergence-free. Our definition of the stress tensor yields a divergence operator that involves the Riemannian structure of both material and spatial manifolds. As a result, the representation of the stress as the curl of a gradient of a scalar function involves both  metrics $\go$ and $\g$ as well: every stress field satisfying \eqref{DivSigma} can be represented as (see \cite{ISF})
\begin{equation}
\sigma^{\mu \nu} = \brk{\frac{1}{\sqrt{|\go|}} \e^{\mu \alpha}} \brk{ \frac{1}{\sqrt{|\g|}} \e^{\nu \beta}} \nabla_{\alpha}\nabla_{\beta} \psi,
\label{SolRep}
\end{equation}
where $\e$ is the Levi-Civita anti-symmetric symbol, $\nabla$ is the covariant derivative with respect to the actual metric, and $|\cdot|$ denotes the determinant.

We call the scalar function $\psi$ the \emph{incompatible stress function} (ISF). It is a  generalization of the Airy stress function for the case of a non-Euclidean reference metric. The representation \eqref{SolRep} involves no approximation, and in particular, does not pertain to a linear theory.

A constitutive relation establishes a relation between the actual metric $\g$ (which determines the strain $u$) and the stress $\sigma$,
\begin{equation}
u = F(\sigma).
\end{equation}
In view of \eqref{SolRep}, a constitutive relation determines equivalently a relation between the ISF and $\g$,
\begin{equation}
\g =\go + 2 F\left(\frac{1}{\sqrt{|\go| \,|\g|}} \e^{\mu \alpha} \e^{\nu \beta} \nabla_{\alpha}\nabla_{\beta} \psi\right).
\label{eq:GenMetric}
\end{equation}

Since $\g$ is an actual metric corresponding to a planar configuration, it must be Euclidean. This yields through \eqref{eq:GenMetric} a geometric constraint on the ISF. We have thus reduced the full elastic problem into that of finding an ISF corresponding to a Euclidean $\g$. If the body is simply connected, the condition that $\g$ be Euclidean reduces to the vanishing of a scalar field---the actual Gaussian curvature, $K_G$. Thus, the elastic problem is reformulated as:
\[
\text{Find $\psi$ such that $K_G=0$ with $\g$ given by \eqref{eq:GenMetric}}.
\]
This reformulation captures both elastic nonlinearity and geometric incompatibility. It is valid for any constitutive relation, as the latter only
affects the relation between $\psi$ and $\g$.

\section{Hookean solids}

The constitutive law  for a Hookean solid is 
\beq
\sigma^{\alpha\beta} = \frac12 \A^{\alpha\beta\gamma\delta}\brk{\g_{\gamma\delta} - \go_{\gamma\delta}},
\label{eq:Hookean}
\eeq
where  $\A$ is the elastic tensor. For a
homogeneous and isotropic material
\[
\A^{\alpha\beta\gamma\delta} = \frac{Y}{1-\nu^2}\brk{(1+\nu)\go^{\alpha\beta}\go^{\gamma\delta}+\nu\go^{\alpha\gamma}\go^{\beta\delta}},
\]
where $Y$ is the Young modulus and $\nu$ is the Poisson ratio.

Inverting \eqref{eq:Hookean}, we express the actual metric in terms of the stress, 
\beq
\g_{\alpha\beta} = \go_{\alpha\beta} + 2 \A_{\alpha\beta\gamma\delta}\sigma^{\gamma\delta},
\label{eq:subhere}
\eeq
where 
\[
\A_{\alpha\beta\gamma\delta} = \frac{1}{Y} \brk{(1+\nu)\go_{\alpha\gamma}\go_{\beta\delta}-\nu\, \go_{\alpha\beta}\go_{\gamma\delta}}.
\]

Substituting the representation  \eqref{SolRep} of the stress into \eqref{eq:subhere},
\begin{equation}
\g_{\mu\nu} = \go_{\mu\nu} + \frac{2 \mathcal{A}_{\mu\nu\alpha\beta}}{\sqrt{|\go|} \sqrt{|\g|}} \varepsilon^{\alpha \gamma} \varepsilon^{\beta \kappa} \nabla_{\gamma}\nabla_{\kappa} \psi.
\label{gsol}
\end{equation}
This expression for $\g$ is implicit as $\g$ appears on the right-hand side both through its determinant as in the definition of the covariant derivative $\nabla$.

\section{Approximate solution for Hookean solids}

The reformulation of the elastic problem derived in the previous section still results in a highly nonlinear problem. In order to devise approximation schemes,
one needs to identify a natural dimensionless parameter that can be used in a perturbative expansion. 
Since our problem results from geometric incompatibility, the expansion parameter is expected to quantify the magnitude of the geometric incompatibility. 

When $\g$ is smooth, every open set of sufficiently small diameter can be embedded in Euclidean space ``almost isometrically". Physically, this means that a small enough sample has a configuration that is almost strain-free. This suggests that for the case of a smooth reference metric, 
a natural expansion parameter is a product of the diameter of the body and a characteristic curvature. 

Let $\eta$ be a small dimensionless  parameter that measures the amount of geometric incompatibility. 
We expand the ISF in powers of $\eta$,
\[
\psi = \eta \, \psi^{(1)} + \eta^2\, \psi^{(2)} + \cdots.
\]
Equation~\eqref{gsol} induces a similar expansion for $\g$,
\[
\g = \go + \eta\, \g^{(1)} + \eta^2\, \g^{(2)} + \cdots,
\]
which in turn induces an expansion for the actual Gaussian curvature,
\[
K_G = \Kbar + \eta\, K_G^{(1)} + \eta^2\, K_G^{(2)} + \cdots.
\]

Since to leading order in $\eta$, $\g$ and $\nabla$ are equal to $\go$ and $\nablago$, it follows from \eqref{gsol} that
\begin{equation}
\g_{\mu\nu} = \go_{\mu\nu} + \frac{2 \eta}{|\go|} \mathcal{A}_{\mu\nu\alpha\beta}  \varepsilon^{\alpha \gamma} \varepsilon^{\beta \kappa} \nablago_{\gamma}\nablago_{\kappa} \psi^{(1)} + O(\eta^2).
\label{FirstOrder}
\end{equation}

The Gaussian curvature is obtained from the metric by
\[
K_G = \frac12 \g^{\alpha \gamma} \g^{\beta \delta} R_{\alpha \beta \gamma \delta},
\]
where $R_{\alpha \beta \gamma \delta}$ is the Riemann curvature tensor \cite{docarmo}.

To leading order in $\eta$, the condition that $K_G=0$ yields an equation for  the leading-order term of the ISF,
\begin{equation}
\begin{split}
Y \Kbar &=
\bar{\Delta} \bar{\Delta} \psi^{(1)} + 
2\Kbar \bar{\Delta} \psi^{(1)}  \\
&+ 
\brk{1+\nu}\go^{\mu \nu} \brk{\partial_\mu \Kbar} \brk{\partial_\nu \psi^{(1)}}.
\end{split}
\label{eq:TheEquation}
\end{equation}
Here $\bar{\Delta}$ is the reference Laplace-Beltrami operator,
\[
\bar{\Delta} f = \frac{1}{\sqrt{|\go|}} \partial_\mu \brk{\sqrt{|\go|} \; \go^{\mu \nu } \partial_\nu f }.
\]
Equation \eqref{eq:TheEquation} together with the boundary conditions determine $\psi^{(1)}$ up to immaterial gauge transformations.
It  is a first-order-approximation, applicable to weak incompatibility. 
For stronger incompatibility higher-order approximations may be needed. 
A general scheme for obtaining higher-order terms of the ISF is presented in
\cite{ISF}.

In the ``compatible" case  $\Kbar=0$. Then \eqref{eq:TheEquation} reduces, as expected, to the biharmonic equation, which is the equation satisfied by the classical Airy stress function. 
If $\Kbar$ is of order $\eta$, \eqref{eq:TheEquation} reduces to
\begin{equation}
\frac{1}{Y} \Delta\Delta \psi^{(1)} = \Kbar.
\label{eq:AppEquation}
\end{equation}
Equation \eqref{eq:AppEquation} is linear both in $\psi^{(1)}$ and in $\Kbar$. Therefore,  the ISF in the presence of multiple defects is the sum of ISF's associated with each defect separately.

Finally, given the solution for the first order ISF $\psi^{(1)}$, the stresses can be calculated to first order in $\eta$ from \eqref{SolRep},
\begin{equation}
\sigma^{\mu \nu} =  \frac{1}{|\go|} \e^{\mu \alpha} \e^{\nu \beta} \nablago_{\alpha}\nablago_{\beta} \psi^{(1)}.
\label{eq:StressRep}
\end{equation}
Since from now on we consider an $O(\eta)$ approximation, we will omit the superscript $(1)$ in the ISF.

\section{Interacting defects}

In this section we calculate the ISF for systems with reference metrics representing  materials with defects, i.e., with reference metrics of the form \eqref{eq:Conformal} and a conformal factor $\vp$ of the form \eqref{eq:mult_phi}. Since $\Kbar$ diverges at the defects loci, it is not clear a priori what constitutes a weak incompatibility. Every localized distribution of Gaussian curvature can be approximated by a smooth distribution. By considering defects whose magnitudes induce length scales that are small compared to the core diameter, we may obtain a reference Gaussian curvature that is everywhere of order $\eta$ and vanishes outside of the defects' core. Under this assumption, the far field stress induced by a singular defect is identical to that induced by the regularlized curvature distribution.

To find the elastic interaction between defects we calculate the total elastic energy stored in the medium at equilibrium in the presence of defects. The energy functional for Hookean solids is
\begin{equation}
E =\frac{1}{2} \int A_{\mu\nu\rho\sigma} \sigma^{\mu\nu}\sigma^{\rho\sigma} \sqrt{|\go|}\, dS.
\label{eq:Energy}
\end{equation}
Substituting \eqref{eq:StressRep},
\begin{equation}
E =\frac{1}{2} \int \frac{1}{|\go|^2} A_{\mu\nu\rho\sigma} \varepsilon^{\mu \alpha} \varepsilon^{\nu \beta} \varepsilon^{\rho \gamma} \varepsilon^{\sigma \delta }\brk{\nablago_{\alpha}\nablago_{\beta} \psi}\brk{\nablago_{\gamma}\nablago_{\delta} \psi} \sqrt{|\go|}\, dS.
\end{equation}
Integrating twice by parts,
\begin{equation}
E =\frac{1}{2} \int \psi \,  \nablago_{\gamma}\nablago_{\delta} \brk{\frac{1}{|\go|^2} A_{\mu\nu\rho\sigma} \varepsilon^{\mu \alpha} \varepsilon^{\nu \beta} \varepsilon^{\rho \gamma} \varepsilon^{\sigma \delta }\brk{\nablago_{\alpha}\nablago_{\beta} \psi} \sqrt{|\go|}}\, dS.
\label{eq:UglyIntegrand}
\end{equation}
Note that $\A$, $\go$ and $\bar{\nabla}$ depend explicit on $\varphi$.
Using once again the fact that $\Kbar$ is small, and expanding the integrand in \eqref{eq:UglyIntegrand} to lowest order in $\eta$, we obtain, 
\begin{equation}
E =\frac{1}{2} \int \psi \,  \brk{\frac{1}{Y} \Delta\Delta \psi}\, dS = \frac{1}{2} \int \psi \,  \Kbar\, dS.
\label{eq:TheResult}
\end{equation}

Equation \eqref{eq:TheResult} bears a strong analogy to electrostatics. 
The Gaussian curvature $\Kbar$ plays the role of an electric charge, whereas the ISF plays the role of an electric potential. 
Like in the electrostatic analog, the self interaction of a charge with its induced potential diverges, and 
should be ignored when considering long range interactions. 
Note also that \eqref{eq:TheResult} is valid both for smooth and singular metrics.

To calculate pairwise interactions between defects, we consider a  material containing  two defects of arbitrary type. Its reference curvature can be represented as sum
\[
\Kbar = \bar{K}_1+\bar{K}_2.
\]
The stress function decomposes into 
\[
\psi = \psi_1 + \psi_2.
\]
The elastic energy without the self-interaction terms is 
\begin{equation}
U=\frac12 \int (\psi_1 \bar{K}_2 + \psi_2 \bar{K}_1)\, dS = \int \psi_1 \bar{K}_2 \,dS,
\label{eq:interactions}
\end{equation}
where the last step follows from integration by parts, as $\int \psi_1\, (\Delta\Delta \psi_2)\, dS = \int  (\Delta\Delta \psi_1)\, \psi_2\, dS$.
Thus,  the pairwise interaction energy between defects depends explicitly on the reference curvatures associated with each defect, and on the ISFs induced by those curvatures.

\begin{figure}[h]
\begin{center}
\includegraphics[width=\columnwidth]{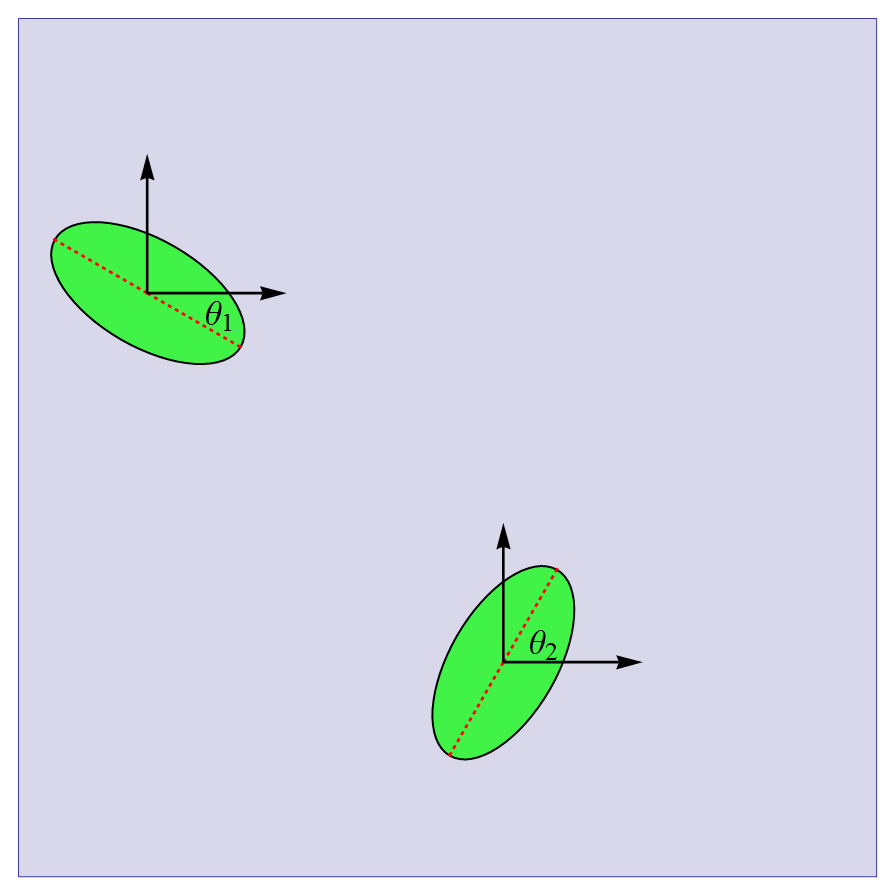}
\end{center}
\caption{Illustration of two quadrupoles induced by a unidirectional contraction of two small regions.}
\label{fig:QQIll}
\end{figure}

Before proceeding to the calculation of interaction energies, let's establish some notations.
Henceforth $\xvec$ denotes the material coordinates and $\hat{\xvec}$ is the corresponding unit vector from the origin to that point.
Given two defects, the vector $\rvec$ denotes their separation and $\hat{\rvec}$ is the corresponding unit vector.
The charge of a dipole is represented by
the vector $\dvec$. The charge of a quadrupole is represented by the tensor
\[
\Qvec = Q \matrixII{\cos 2\theta}{\sin 2\theta}{\sin 2\theta}{- \cos 2\theta},
\]
where $\theta$ is the quadrupole orientation relative to the $x$-axis (\figref{fig:QQIll}).

Another type of singularity is point defects, corresponding to an isotropic expansion/contraction of a disc. We denote the charge of a point defect by $P$; it is positive for expansion and negative for contraction.

In Table~\ref{tab:table1} we list the reference curvatures for the most ubiquitous types of defects. The corresponding ISF is obtained by solving \eqref{eq:AppEquation}; it is determined modulo immaterial solutions of the homogeneous biharmonic equation. The function $\Cof{\cdot}$ appearing in the case of an external field is the cofactor of a matrix.

\begin{table}[h]
\begin{ruledtabular}
\begin{tabular}{lcc}
Type   &  $\bar{K}$ & $\psi$  \\ \hline
Dipole & 
$ \dvec\cdot\nabla\delta(\xvec)$
&
$(Y/4\pi)\,(\dvec\cdot \xvec) \ln |\xvec|$ 
\\ 
Quadrupole & 
$\frac14 (\nabla^T\cdot \Qvec\cdot \nabla) \delta(\xvec-\rvec)$ 
&
$(Y/16\pi)\, ({\hat{\xvec}^T}\cdot\Qvec\cdot\hat{\xvec} )$ 
\\
Point & 
$- 2 P \,\Delta \delta(\xvec)$ 
&
$-(Y P \,/2\pi) \ln|\xvec|$
\\
External & 
0
&
$\frac12 (\xvec^T \cdot \Cof{\sigma} \cdot \xvec)$  
\\
\end{tabular}
\end{ruledtabular}
\caption{Reference curvatures and ISF's for various types of defects.
The bottom line corresponds to an external stress field $\sigma$.}
\label{tab:table1}
\end{table}

To obtain the interaction energy between two defects one needs to substitute in \eqref{eq:interactions} the ISF associated with one defect and the reference curvature associated with the other. In \tabref{tab:table2} we list the interaction energy between  pairs of defects and between defects and an external load. For brevity we denote $\langle A \rangle \equiv \hat{\rvec}^T \cdot A \cdot \hat{\rvec}$

\begin{table}[h]
\begin{ruledtabular}
\begin{tabular}{lc}
Defect types & $U$\\ \hline
 Dipole-Dipole &  $-(Y/4\pi)\, \brk{\dvec_1 \cdot \dvec_2  \, \ln |\rvec| + \brk{\dvec_1 \cdot \hat{\rvec}} \brk{\dvec_2 \cdot \hat{\rvec}}}$\\ 
 Quad-Quad & $(Y/16\pi r^2)\, \brk{2 \langle\Qvec_1\rangle \langle\Qvec_2\rangle - \langle\Qvec_1 \Qvec_2\rangle}$ \\
 Dipole-Quad& $(Y/8\pi r)\, \brk{\hat{\rvec}^T \cdot \Qvec \cdot \dvec - \langle\Qvec\rangle(\hat{\rvec}^T \cdot \dvec )}$\\
 Point-Dipole& $(Y P \, / \pi r) \,(\hat{\rvec}^T \cdot \dvec)$\\
 Point-Quad&  $(Y  P\, /2 \pi r^2) \langle\Qvec\rangle$\\
 Point-Point&  $0$\\
 External-Dipole& $0$\\
 External-Quad& $-\frac{Q}{4} (\sigma_0^{xx} - \sigma_0^{yy})\cos 2\theta - \frac{Q}{2} \sigma_0^{xy} \sin 2\theta$\\
 External-Point& $-2 P  (\sigma_0^{xx}+\sigma_0^{yy})$\\
\end{tabular}
\end{ruledtabular}
\caption{Interactions energies between pairs of defects and between a defect and an external stress field.}
\label{tab:table2} 
\end{table}

The interaction energy between a pair of dipoles (i.e., dislocations) agrees with the classical result of Nabarro \cite{Nabarro1952}. It should be noted, however, that in contrast with Nabarro's result, our results are limited to effectively 2D systems, that is all vectors in Table II are two-dimensional.
Moreover, as known for linearized models of isotropic Hookean solids, pairs of point defects do not interact with each other \cite{Ardell1966}. It should be emphasized that such results are model dependent---point defects, for example, might interact under a different constitutive relation.

\section{Applications}


\subsection{Localized plastic deformations}


Consider an amorphous material subject to an external shear stress. 
If the response is purely elastic, elastic deformations result in stored elastic energy. If, however, plastic deformations are allowed (in our language, the reference metric $\go$ can change), then the stored elastic energy can be reduced. In fact, the elastic energy can be totally eliminated by forming 
 a spatially uniform distribution of defects of the quadrupolar type.

Indeed,
the stress field induced by a single quadrupole located at $(x_0,y_0)$ with orientation $\theta = \pi/4$ is
\begin{equation}
\begin{aligned}
\sigma^{xx} &= \partial_{yy}\psi_Q(x-x_0, y-y_0) \\
\sigma^{yy} &= \partial_{xx}\psi_Q(x-x_0, y-y_0) \\
\sigma^{xy} &= -\partial_{xy}\psi_Q(x-x_0, y-y_0),
\end{aligned}
\label{eq:single_Q}
\end{equation}
where $\psi_Q$ is given in \tabref{tab:table2}.
The stress field induced by a uniform distribution of identical quadrupoles is obtained by integrating \eqref{eq:single_Q} with respect to $x_0$ and $y_0$. For a quadrupole charge density $q$ we obtain,
\begin{equation}
\begin{aligned}
\sigma^{xx} &= 0 \\
\sigma^{yy} &= 0 \\
\sigma^{xy} &= -\frac{Y q}{4\pi},
\end{aligned}
\end{equation}
which is a pure shear stress.
Thus, given an external shear stress $\sigma^0$, a uniform distribution of quadrupoles  with orientation $\theta = \pi/4$ and charge density
\[
q = \frac{4\pi \sigma^0}{Y}
\]
results in a stress-free state, indicating that the formation of quadrupoles is an energetically-favorable response to external shear.

A material that does not allow for plastic deformations is an elastic solid, whereas a material that allows for continuous plastic deformations is liquid-like. Plasticity models assume that plastic deformations only occur beyond a stress threshold \cite{Kobayashi2012}. 
Thus, a defect will only form at a location in which the stress exceeds the threshold. The formation of a new defect changes the ambient stress field. While the total elastic energy is always reduced by the formation of a new defect, local stresses may grow, leading to the formation of new defects.

Consider a pair of quadrupoles in an external shear stress field.
The total interaction energy is
\begin{equation}
\begin{split}
U &= -\frac{\sigma_0 Q_1}{2} \sin 2\theta_1-\frac{\sigma_0 Q_2}{2} \sin 2\theta_2\\ 
&+ \frac{Y Q_1 Q_2}{16 \pi r^2} \cos(2\theta_1 + 2\theta_2-4\phi),
\end{split}
\label{eq:QuadInt}
\end{equation}
where $Q_i$ are the quadrupoles charges, $\theta_i$ are their orientations relative to the shear direction, $\sigma_0$ is the external shear stress, and $\rvec = ( r \cos \phi, r \sin \phi)$ is their spatial separation vector. 

The  first two terms represent the interaction of the quadrupoles with the external shear stress. The fact that  they can be negative indicates that it is energetically-favorable to generate quadrupoles as a response to external shear.  The minimum energy state for the first two terms is obtained for $\theta_1= \theta_2 = \pi/4$.
The third term is the quadrupole-quadrupole (Q-Q) interaction. It can be minimized simultaneously with the first two terms by taking $\phi = 0$. This implies that in the minimum energy state the line separating the quadrupoles is parallel to the direction of the shear whereas the principal axis of the quadrupolar moments is at an angle of $\pi/4$ with the shear direction. The same considerations remain valid with $N$ quadrupoles: a state of minimal energy is obtained when all the quadrupoles lie on a line parallel to the direction of the shear and are oriented at an angle of $\pi/4$.

This result is in agreement with Dasgupta et al. \cite{Procaccia}, which showed, both analytically and numerically, that in the limit of large external loads, localized plastic deformations form along a line parallel to the external shear. Moreover, it was suggested that the formation of a linear array of quadrupoles initiates the failure process of the material.

Another known property of solids is the avalanche-like behavior of plasticity: the nucleation of a small number of localized plastic deformations initiates a rapid formation of more localized plastic deformations, leading eventually to failure. To explain this phenomenon we assume that a localized plastic deformation will form at a point only if the local stress exceeds a critical threshold $\sigma_{crit}$ (see \cite{Kobayashi2012}).

The stress field induced by a single quadrupole is
\begin{equation}
\begin{aligned}
\begin{split}
\sigma^{xx} &= -\frac{3 Q \sin\brk{4\theta}}{2\pi r^4} \\
\sigma^{yy} &= \frac{3 Q \sin\brk{4\theta}}{2\pi r^4} \\
\sigma^{xy} &= \frac{3 Q \cos\brk{4\theta}}{2\pi r^4}.
\end{split}
\end{aligned}
\end{equation}
This stress enhances the external shear stress along the $\theta=0,\pi/2$ directions, and reduces the shear stress along the $\theta=\pm\pi/4$ directions. Once several quadrupoles have formed along the $\theta=0$ axis, which as we saw is the energetically-favorable configuration, their presence reduces total shear stress almost everywhere, but it builds up an even stronger shear stress on the line that connects them, thus increasing the probability of defect formation along that same line. This observation might explain the avalanche of defect formation preceding failure.

\subsection{Interacting active cells}

Living cells adhering to a substrate exert force on the substrate when undergoing conformal changes \cite{Balaban2001,Kineret2013}.
If the substrate is elastic, it mediates mechanical interactions between deforming cells. There exists an extensive literature on this subject (see e.g., \cite{Safran2013}).

\begin{figure*}
\begin{center}
\includegraphics[width=6in]{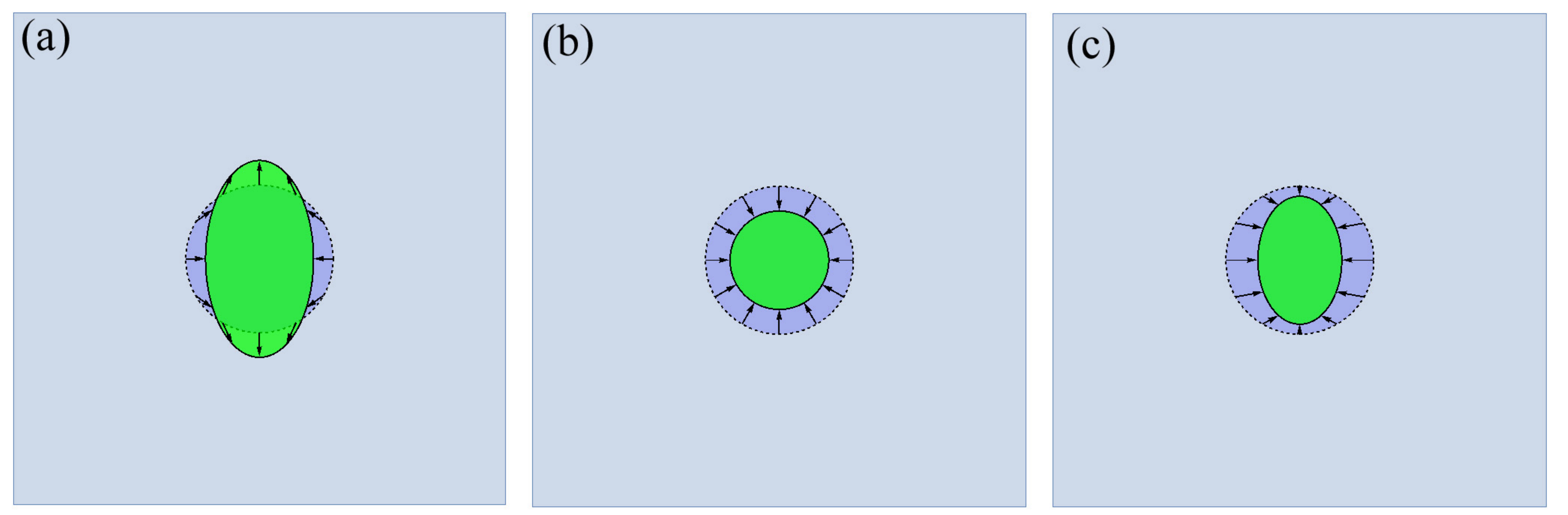}
\end{center}
\caption{A schematic illustration of modes of deformation of a contracting cell. (a) A deformation that almost preserves area, but significantly changes the eccentricity. In this case $P\ll Q.$ (b) A deformation that almost preserves the eccentricity but significantly changes the area. In this case $Q\ll P$. Contraction and expansion correspond to negative and positive values of $P$, respectively. (c) A deformation that changes both eccentricity and area, that is $P \approx Q$.}
\label{fig:ContCells}
\end{figure*}

Consider a single cell adhered to a 2D elastic substrate. When the cell deforms, it causes the substrate under its ``footprint" to deform as well, thus constituting a local source of stress. Assuming the unperturbed substrate to be Euclidean, such a source of stress can be modeled as a singularity in the reference curvature. 

If the size of a cell is significantly smaller than inter-cellular separations, one can approximate the reference metric by its lowest-order multipoles. Since the cells' conformational changes induce local metric perturbations, there are neither monopoles not dipole charges, hence the lowest-order multipoles correspond to quadrupoles  and point defects. Under these assumptions, the reference curvature induced by a single cell located at the origin is of the form
\[
K_\text{Cell} = \frac{1}{4} (\nabla^{T} \cdot \Dvec \cdot \nabla) \delta(\xvec),
\]
where
\[
\Dvec = 2 P \matrixII{1}{0}{0}{1}+ Q \matrixII{\cos 2\theta}{\sin 2\theta}{\sin 2\theta}{-\cos 2\theta}.
\]
Here $P$ is the charge of a point defect, induced by a change in the area of the cell's footprint. The parameter $Q$ is the quadrupole charge associated with the eccentricity of the cell's deformation. 

Consider two deforming cells adhering to a Hookean substrate. By \tabref{tab:table2}, 
the interaction energy between the cells is 

\begin{equation}
\begin{aligned}
\begin{split}
U_\text{Cells} &= \frac{Y Q_1 Q_2}{16 \pi R^2} \left(\cos(2 \xi_1 + 2\xi_2)\phantom{\frac{1}{1}}\right. \\&\left.\phantom{\frac{1}{1}}+ 2 \rho_1 \cos(2 \xi_2)+ 2 \rho_2 \cos(2 \xi_1)\right) ,
\end{split}
\end{aligned}
\label{eq:CellInt}
\end{equation}

where $P_1,P_2,Q_1,Q_2$ are the point charges and quadrupole charges associated with the deformations, $\rho_i = -P_i/Q_i$ are dimensionless measures of the isotropy of the deformations,
$R$ is the distance between the cells, $\xi_1$ and $\xi_2$ are the orientations of the two quadrupoles with respect to the line separating them.

Different behaviors are expected for different values of the deformation parameters $\rho_1$ and $\rho_2$. 
%
%
If both cells deform almost isotropically, $1 \ll |\rho_i|$, then the interaction energy is dominated by the last two terms in \eqref{eq:CellInt} ($Q-P$ interactions). 
If the cells contract ($0<\rho_i$), then the principal axes of deformation are $\xi_{1,2} = \pi/2$, i.e.,
perpendicular to the line separating the cells.
If the cells expand ($\rho_i<0$) then the principal axes of deformation are $\xi_{1,2} = 0$, i.e.,
parallel to the line separating the cells.

If both cells deform almost unidirectionally, $|\rho| \ll 1$,  then the interaction energy is dominated by the first term in \eqref{eq:CellInt} ($Q-Q$ interaction). Then the preferred orientations satisfy $\xi_1+\xi_2 = \pi/2$, which is a degenerate solution. Considering also the $Q-P$ interactions removes this degeneracy: for $\rho_j <\rho_i$ the energy minimizing orientations are $\xi_i = 0$ and $\xi_j = \pi/2$.


The difference between the limiting regimes suggests the existence of a phase diagram for cells' orientations.  In \figref{fig:MinTheta1} we show the energy-minimizing value of $\xi_1$ as function of $\rho_1$ and $\rho_2$.
{The corresponding phase diagram is shown in \figref{fig:Bifurcation}. 
\begin{figure}[h]
\begin{center}
\includegraphics[width=\columnwidth]{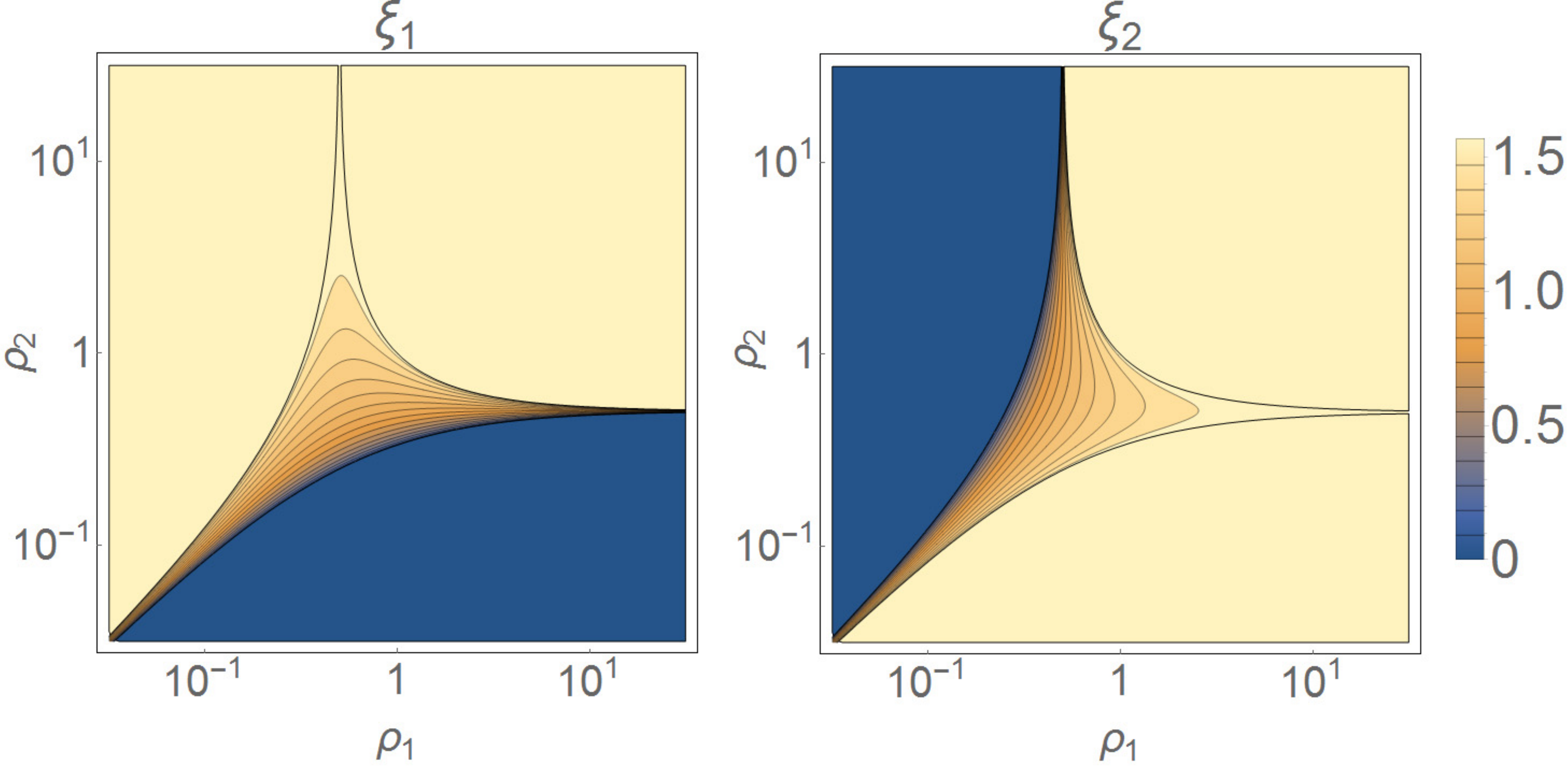}
\end{center}
\caption{Contour plots of the energy minimizing cell orientations $\xi_1$ and $\xi_2$ as functions of $\rho_1$ and $\rho_2$.}
\label{fig:MinTheta1}
\end{figure}

\begin{figure}[h]
\begin{center}
\includegraphics[width=\columnwidth]{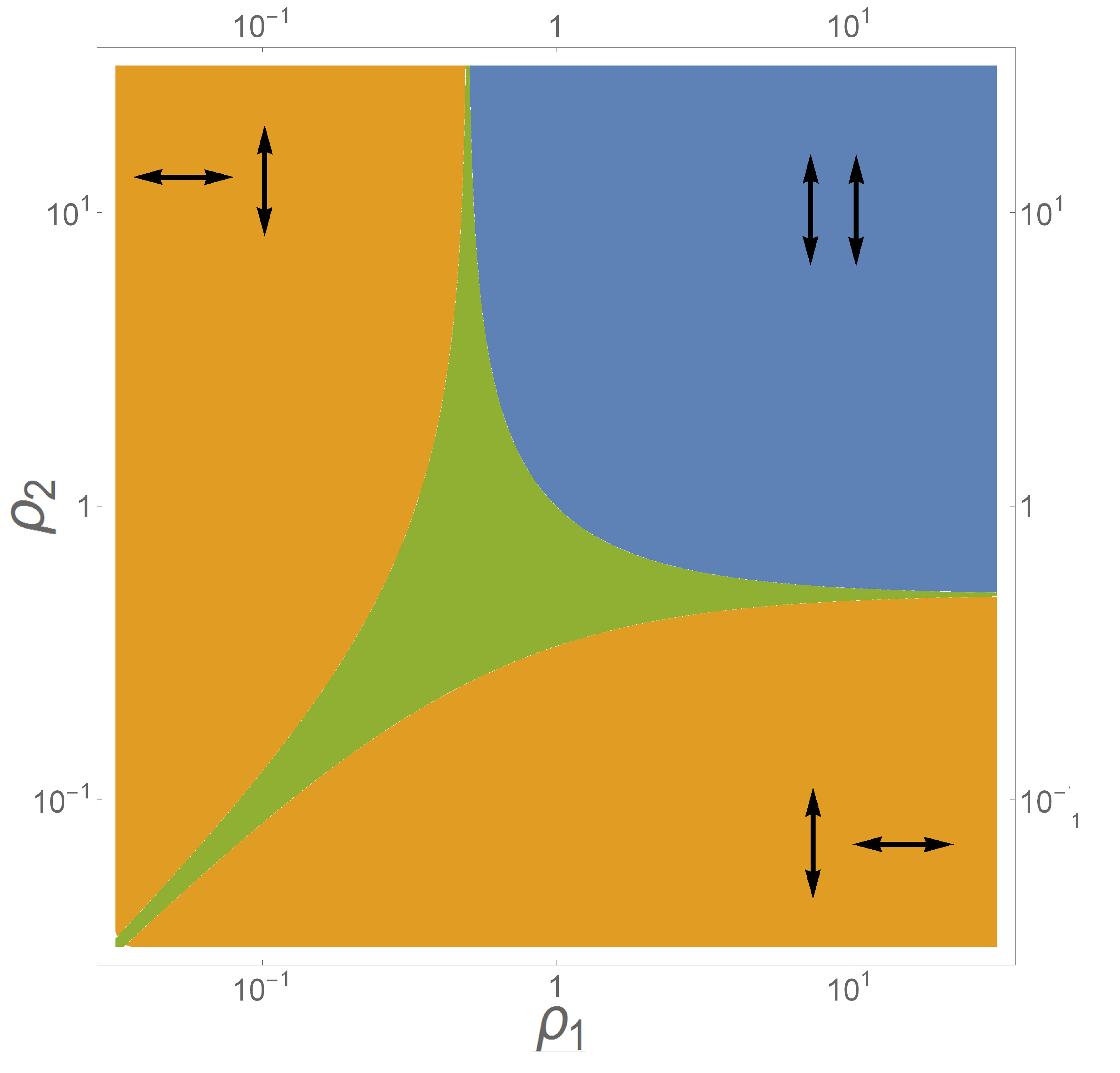}
\end{center}
\caption{A phase diagram for the interaction between two deforming cells adhering to a Hookean elastic substrate, as a function of the deformations isotropies $\rho_1$ and $\rho_2$. For $\rho$'s in the blue region, the cells are parallel to each other, and perpendicular to the line separating them. For $\rho$'s in the Yellow regions the cells are perpendicular to each other, where one is parallel to the line separating them. The green region does not correspond to a well defined phase. The orientations vary smoothly inside this region, and continuously at the phase separating curves.}
\label{fig:Bifurcation}
\end{figure}

In the blue domain, which corresponds to the case where both cells deform almost isotropically, the  principal axes of deformation of the two cells are parallel to each other, and perpendicular to the line connecting them. 
The yellow regions, which corresponds to the case where the two cells deform differently,  displays one quadrupole parallel to the line separating them and the other perpendicular to it.
The green region does not corresponds to a well defined interaction phase.

\section{Discussion}

Localized sources of stress appear in a variety of physical systems. Examples of such stress sources are defects in crystalline material, local plastic deformations in amorphous materials, and living contracting cells attached to elastic substrate. Despite the similarity between these systems, their study in the existing literature involves different approaches and methodologies. Defects, for example, are usually described as deviations from an ordered state, whereas contracting cells are commonly modeled as localized forces. 


In the current work we present a unified framework for local sources of stress in effectively 2D elastic media. Sources of stress are viewed as singularities in a reference curvature field, which can be interpreted as "elastic charges".
This approach, which was demonstrated for Hookean solids, can be applied to other constitutive laws. 

While the resulting set of equations is implicit and nonlinear, we derived a perturbative approximation, which in principle, can be carried to any desired order. 

The first-order problem has a structure reminiscent of electrostatics. Each defect has a ``charge" which induces a specific ``potential". The elastic energy density is the product of the local charge and  potential. This formulation opens the way to handle many different elastic problems in a surprisingly simple way. The applicability of the formulation is not limited to the problems solved in this paper. We believe our approach to be applicable to many other settings, such as multiple interacting strain sources and problems that involve non-simply-connected bodies.

A potential extension of this work is interactions between 3D defects. While a 2D defect is characterized by a singular reference Gaussian curvature, a 3D defect should be characterized by a singular Riemann curvature tensor. The analysis of 3D defects, however, requires the extension of the current tools for solving elastic problems in 3D geometrically-incompatible elastic materials. 

\section{Acknowledgments}
M.M.and E.S. were supported by the Israel- US Binational Foundation (Grant No. 2008432) and by the European Research Council SoftGrowth project. M.M and R. K. was supported by the Israel-US Binational Foundation (Grant No. 2010129) and by the Israel Science Foundation (Grant No. 661/13).


\begin{thebibliography}{10}
\bibitem{Taylor1}
Taylor, G. I. The mechanism of plastic deformation of crystals. parts 1-3. Proceedings of the Royal Society of London A: Mathematical, Physical and Engineering Sciences 145, 362– 415 (1934).

\bibitem{Moshe2015}
Moshe, M., Levin, I., Aharoni, H., Kupferman, R. \& Sharon, E. Proceedings of the National Academy of Sciences 112, 10873–10878 (2015).

\bibitem{Procaccia2} Dasgupta, R., Hentschel, H. G. E. \& Procaccia, I. Microscopic mechanism of shear bands in amorphous solids. Phys. Rev. Lett. 109, 255502 (2012).

\bibitem{Nabarro1952} Nabarro, F. R. N. Mathematical theory of stationary dis- locations. Advances in Physics 1, 269–394 (1952).

\bibitem{Michel1980} Michel, B. Interaction between a point defect and an edge dislocation near solid surfaces. physica status solidi (b) 99, K1–K4 (1980).

\bibitem{ISF} Moshe, M., Sharon, E. \& Kupferman, R. The plane stress state of residually stressed bodies: a stress function approach. arXiv preprint arXiv:1409.6594 (2014).

\bibitem{Balaban2001} Balaban, N. Q. et al. Force and focal adhesion assembly: a close relationship studied using elastic micropatterned substrates. Nature cell biology 3, 466–472 (2001).

\bibitem{Kineret2013} Shah, E. A. \& Keren, K. Mechanical forces and feedbacks in cell motility. Current opinion in cell biology 25, 550–557 (2013).

\bibitem{Safran2013} Schwarz, U. S. \& Safran, S. A. Physics of adherent cells. Reviews of Modern Physics 85, 1327 (2013).

\bibitem{EfiJMPS} Efrati, E., Sharon, E. \& Kupferman, R. Elastic theory of unconstrained non-euclidean plates. Journal of the Mechanics and Physics of Solids 57, 762–775 (2009).

\bibitem{Neff2015} Neff, P., Eidel, B. \& Martin, R. J. Geometry of logarithmic strain measures in solid mechanics. arXiv preprint arXiv:1505.02203 (2015).

\bibitem{Kondo} Kazuo, K. On the analytical and physical foundations of the theory of dislocations and yielding by the differential geometry of continua. International Journal of Engineering
Science 2, 219–251 (1964).

\bibitem{Bilby} Bilby, B. A., Bullough, R. \& Smith, E. Continuous Distributions of Dislocations: A New Application of the Methods of Non-Riemannian Geometry. Proceedings of the Royal Society A: Mathematical, Physical and Engineering Sciences 231, 263–273 (1955).

\bibitem{kroner} Kroner, E. Physique des De ́fauts, Continuum theory of Defects (1980).

\bibitem{Wang1968} Wang, C.-C. On the Geometric Structures of Simple Bod- ies, a Mathematical Foundation for the Theory of Continuous Distributions of Dislocations. 87–148 (Springer Berlin Heidelberg, 1968).

\bibitem{doCarmoBaby} DoCarmo, M. P. Differential Geometry Of Curves And Surfaces (1976).

\bibitem{liouville1853equation} Liouville, J. Journal de mathematiques pures et appliquees 71–72 (1853).

\bibitem{docarmo} do Carmo Valero, Manfredo Perdigao. Riemannian geometry. 1992..

\bibitem{Ardell1966} Ardell, A. \& Nicholson, R. On the modulated structure of aged ni-al alloys: with an appendix on the elastic interaction between inclusions by JD Eshelby Cavendish laboratory, University of Cambridge, England. Acta metallurgica 14, 1295–1309 (1966).

\bibitem{Kobayashi2012} Kobayashi, T. Strength and toughness of materials (Springer Science \& Business Media, 2012).

\bibitem{Procaccia} Dasgupta, R., Karmakar, S. \& Procaccia, I. Universality of the plastic instability in strained amorphous solids. Phys. Rev. Lett. 108, 075701 (2012).


\end{thebibliography}

\end{document}